# THE ACCURACY OF CELL-BASED DYNAMIC TRAFFIC ASSIGNMENT: IMPACT OF SIGNAL CONTROL ON SYSTEM OPTIMALITY

T. ISLAM [a], H. L. VU [a], M. PANDA [a], N. HOANG [a], D. NGODUY [b]
[a]*Faculty of Science Engineering and Technology,*
*Swinburne University of Technology, Victoria, Australia*
Email: tislam, hvu, mpanda, hhoang@swin.edu.au
[b]*Institute for Transport Studies*, *University of Leeds*, *United Kingdom*
Email: d.ngoduy@leeds.ac.uk

**ABSTRACT**

Dynamic Traffic Assignment (DTA) provides an approach to determine the optimal path and/or departure time based on the transportation network characteristics and user behavior (e.g., selfish or social). In the literature, most of the contributions study DTA problems without including traffic signal control in the framework. The few contributions that report signal control models are either mixed-integer or non-linear formulations and computationally intractable. The only continuous linear signal control method presented in the literature is the Cycle-length Same as Discrete Time-interval (CSDT) control scheme. This model entails trade-off between cycle-length and cell-length. Furthermore, this approach compromises accuracy and usability of the solutions.

In this study, we propose a novel signal control model namely, Signal Control with Realistic Cycle-length (SCRC) which overcomes the trade-off between cycle-length and cell-length and strikes a balance between complexity and accuracy. The underlying idea of this model is to use a different time scale for the cycle-length. This time scale can be set to any multiple of the time slot of the Dynamic Network Loading (DNL) model (e.g. CTM, TTM, and LTM) and enables us to set realistic lengths for the signal control cycles. Results show, the SCRC model not only attains accuracy comparable to the CSDT model but also more resilient against extreme traffic conditions. Furthermore, the presented approach substantially reduces computational complexity and can attain solution faster.

## 1. INTRODUCTION

Dynamic Traffic Assignment (DTA) is an approach that describes the interplay between optimal route choice, and traffic flows with time and cost. When a DTA problem is solved from the system operator's point of view, it is called the System Optimal Dynamic Traffic Assignment (SO-DTA). However, the SO-DTA framework is incomplete without Signal Control (SC), as the SC determines how the traffic traverses the network.

Dynamic Network Loading (DNL) models specify how traffic propagates over a given network in space and time. The Cell Transmission Model is a widely accepted DNL model. The CTM is the discrete analogue of the hydrodynamic flow-density differential equations. Compared to other macroscopic DNL models (e.g. METANET, DYNALOAD, MCKW), the CTM provides relatively realistic details about queue formation, propagation and dissipation of congestion through kinematic waves (Nie and Zhang, 2008). In CTM, the road segments are divided into a number of homogeneous units called cells. The accuracy of CTM increases with the shorter duration of the time slots (Sun and Bayen, 2008). However, shorter slots impose larger number of cells per link that leads to an increased computational complexity and time.

The CTM was first proposed by Daganzo (1994). Lately, Daganzo (1995) extended the model by introducing three-legged junctions which can represent more complex networks. However, both of the presented approaches are non-linear in nature (Eq. (4)) and impose very high complexity and intractable computational burden to be used in an optimization framework. Based on Daganzo's work, Ziliaskopoulos (2000) has relaxed the non-linear difference equations to a simple set of linear relationships. Due to the linearity property of this formulation, it is tractable for realistic size network. Waller and Ziliaskopoulos (2001; 2006) address the uncertainty of traveller demand for single destination network of the SO-DTA problem. However, none of these contributions include traffic signal control.

Several network-wide system optimal traffic assignment and mixed integer signal optimization formulations are presented in the literature (Aziz and Ukkusuri, 2012; Beard and Ziliaskopoulos, 2006; Lo, 1999; Zhang, Yin and Lou, 2010). The mixed-integer problem arises as each of the time units is assigned to a certain phase and traffic belonging to all other phases has to wait till the respective phase gets the green time. A continuous but non-linear signal control formulation is presented by Pohlmann and Friedrich (2010). However, the mixed-integer or non-linear formulation has very high complexity and may not attain solution for large networks.

The majority of the SC models presented in the literature are mixed-integer or non-linear in nature and impose enormous computational burden. Only one linear-continuous signal control model was found in the literature (Ukkusuri, Ramadurai and Patil, 2010). Nevertheless, in this approach the cycle-length can only be set equal to the time slots of the DNL model. As a result, for setting up a practically large cycle-length, this model entails an impractically large time slot, hence, very large cell length. As a consequence, this approach would compromise accuracy of the solution.

To increase the accuracy of the solution, one requires shorter time slot. On the other hand, shorter time slot restricts cycle-length to be impractically smaller in duration and increases number of cells per link, hence, increased complexity of the problem. Therefore, the time scale of any discrete DNL model enforces trade-off between accuracy and complexity. In this research, we investigate how the signal control cycle-length can be increased to a practically large extent for a reasonably short time slot. The contributions of this research are as follows:

- *We present a linear-continuous System Optimal Dynamic Traffic Assignment with Signal Control (SO-DTA-SC) formulation. This framework overcomes the cell-length cycle-length trade-off by introducing an extended cycle-length concept, namely, Signal Control with Realistic Cycle-length (SCRC). The SCRC model includes a novel idea of using different time scales to achieve large cycle-length where one traffic signal cycle slot is equal to the several time units of the DNL model.*
- *This study quantifies the trade-off between performance, accuracy, and complexity as a function of the cycle-length.*

To the best of authors' knowledge, this approach is not yet presented in the literature and is a novel contribution in the field. The presented model not only resolves the cell-length cycle-length trade-off but also reduces the complexity of the solution framework by a reasonable amount. Furthermore, this model can be applied to any discrete DNL model such as TTM (Balijepalli, Ngoduy and Watling, 2013) to formulate an SO-DTA-SC framework that can solve any realistic size network. The rest of the paper is organized as follows. Section 2 outlines the linear-continuous SO-DTA-SC formulation; Section 3 presents results and analysis of the solutions. Section 4 concludes this paper by summarizing the key outcomes of this study.

## 2. MODEL FORMULATION

In this section, we develop the Signal Control with Realistic Cycle-length (SCRC) model. Standard SO-DTA framework suck as Ziliaskopoulos' (2000) formulation has both the holding at the source and vehicle holding-back problem. To resolve holding at the source problem, we propose two alternative objectives. For finding the ground truth both the Cycle-length Same as Discrete Time-interval (CSDT) and Mixed Integer Signal Control (MISC) models are also discussed. A list of notation used throughout this paper is presented below.

$v$ = free flow speed, $w$ = backward propagation speed, $\delta_i^t$ = ratio of $v/w$, $\mathcal{C}$ = set of all cells,

$\mathcal{E}$ = set of all connectors, $\mathcal{T}$ = set of all discrete time intervals, $\tau$ = length of each time slot, $\mathcal{C}_D$ = set of diverging cells, $\mathcal{C}_M$ = set of merge cells, $\mathcal{C}_O$ = set of ordinary cells, $\mathcal{C}_S$ = set of sink cells, $\mathcal{C}_R$ = set of source cells, $\mathcal{E}_R$ = set of source cell connectors, $\mathcal{E}_O$ = set of ordinary cell connectors, $\mathcal{E}_M$ = set of merge cell connectors, $\mathcal{E}_D$ = set of diverging cell connectors, $\mathcal{C}_I$ = set of intersection cells, $d_i^t$ = demand at the source cell $i$ at interval $t$, $x_i^t$ = occupancy of cell $i$ at time slot $t$, $N_i^t$ = maximum number of vehicles in cell cell $i$ at time slot $t$, $y_{ij}^t$ = flow from cell $i$ to $j$ at interval $t$, $Q_i^t$ = maximum flow into/out of cell $i$, $\Gamma(i)$ = set of successor cells to cell $i$, $\Gamma^{-1}(i)$ = set of predecessor cells, $\mathcal{E}_{IM}$ = set of intersection merge connectors, $c$ = cycle start index, $G_i^c$ = green split for the $\mathcal{C}_I$ cell $i$ at cycle $c$, $p$ = phase index, $P$ = set of all the phases, $\kappa_{ij}^t = 1$, if the movement from cell $i$ is allowed, $\xi_p^t = 1$, if the phase is active, $\sigma_{ij}$ = set of phases that use cell $i \in \mathcal{C}_{IM}$ when $j \in \Gamma^{-1}(i)$, $G_{min}$ = minimum green split, $\mathcal{I}$ = set of all intersections.

## 2.1 Objective Functions

The proposed SO-DTA formulation presented by Ziliaskopoulos (2000) has both the holding-at-the source and vehicle holding-back problem for the objective of Eq. (1). However, the holding at the source problem can be resolved by adding extra cost over the source cells or adding benefit to the objective for increasing flow out of the source cell. In this section, we propose different objectives that can successfully resolve holding at the source problem.

$$Min \sum_{\forall t \in \mathcal{T}} \sum_{\forall i \in \mathcal{C}/\mathcal{C}_S} x_i^t \qquad (1)$$

### 2.1.1 Flow Benefit Objective

An objective that can resolve holding at the source is presented in Eq. (2). The first term implies minimization of overall system-wide occupancy. The second term in the objective is the benefit for remaining time slots $T - t$ if one increases the flow by one unit at time $t$ from any cell $i$. As a result, this objective forces earliest departure of demand from the source cells and tries to attain earliest arrival pattern at the sink cells.

$$Min \sum_{\forall t \in \mathcal{T}} \sum_{\forall i \in \mathcal{C}/\mathcal{C}_S} x_i^t - \sum_{\forall t \in \mathcal{T}} \sum_{\forall (i,j) \in \mathcal{E}} \alpha \cdot y_{ij}^t \cdot (T+1-t) \qquad (2)$$

### 2.1.2 Double Cost on Source Occupancy (DCS)

This objective also resolves holding at the source problem by applying extra cost over the source cell. As a result, this extra cost forces the solver to inject more vehicles into the network. The objective presented in Eq. (3) includes both the occupancy minimization SO objective and extra penalty for holding at the source cell.

$$Min \sum_{\forall t \in \mathcal{T}} \sum_{\forall i \in \mathcal{C}/\mathcal{C}_S} x_i^t + \sum_{\forall t \in \mathcal{T}} \sum_{\forall i \in \mathcal{C}_R} x_i^t \qquad (3)$$

## 2.2 Constraints

The vehicles propagations over the links are based on the linearized CTM approach of Ziliaskopoulos (2000). Due to the linearity property of this formulation, it is tractable for realistic size networks and used as a base element in this study. The linear conservation equations for the flow-occupancy balance relation for different cells can be found in Ziliaskopoulos (2000). The non-linear CTM flow propagation rule introduced by Daganzo (1995; 1996) is outlined in Eq. (4) where the min operator levies the non-linearity.

$$y_{ki}^t = min\{x_k^t, min[Q_i^t, Q_k^t], \delta_i^t(N_i^t - x_i^t)\} \qquad (4)$$

The above model was relaxed and linearized into number of constraints by Ziliaskopoulos. The constraints are presented in Eq. (5) to Eq. (9). Reader should refer to Ziliaskopoulos (2000) for details about these models.

$$y_{ij}^t - x_i^t \leq 0, \ y_{ij}^t \leq Q_i^t, \ \forall (i,j) \in \mathcal{E}_{R \cup O}, \ j \in \Gamma(i), \ \forall t \in \mathcal{T}. \quad (5)$$

$$y_{ij}^t \leq Q_j^t, \ y_{ij}^t \leq \delta_j^t(N_j^t - x_j^t), \ \forall (i,j) \in \mathcal{E}_{R \cup O \cup D}, \ j \in \Gamma(i), \ \forall t \in \mathcal{T}. \quad (6)$$

$$\sum_{j \in \Gamma(i)} y_{ij}^t - x_i^t \leq 0, \ \sum_{j \in \Gamma(i)} y_{ij}^t \leq Q_i^t, \ \forall (i,j) \in \mathcal{E}_D, \ j \in \Gamma(i), \ \forall t \in \mathcal{T}. \quad (7)$$

$$y_{ki}^t \leq Q_k^t, \ y_{ki}^t - x_k^t \leq 0, \ \forall (k,i) \in \mathcal{E}_{M \cup S}, \ k \in \Gamma^{-1}(i), \ \forall t \in \mathcal{T}. \quad (8)$$

$$\sum_{k \in \Gamma^{-1}(i)} y_{ki}^t \leq \delta_i^t(N_i^t - x_i^t), \ \sum_{k \in \Gamma^{-1}(i)} y_{ki}^t \leq Q_i^t, \ \forall (k,i) \in \mathcal{E}_M, \ k \in \Gamma^{-1}(i), \ \forall t \in \mathcal{T}. \quad (9)$$

## 2.3 Signal Control Models

In this section, constraints related to the signal control models are presented. Along with the novel Signal Control with Realistic Cycle-length (SCRC) model two other models are also discussed.

### 2.3.1 The Proposed Model

The proposed Signal Control with Realistic Cycle-length (SCRC) model enables us to set signal control cycle-length beyond the time slot length of any discrete Dynamic Network Loading model (e.g CTM, TTM). Furthermore, the presented model has linear-continuous formulation and easier to solve for any realistic size network.

The underlying idea of the signal control model is that it computes the green split at the each of the cycle-start-index and applies the green split for the rest of the time slots within that cycle. The cycle-length $m$ can be set to any multiple of the discrete time interval. In Eq. (10), we can see the signal control split $G_i^c$ are applied to the intersection cells $\mathcal{C}_I$ as reduced maximum flow capacity of $Q_i^t$. In the equation, cycle start index is $c = \left\lfloor \frac{t-1}{m} + 1 \right\rfloor$ where $1 \leq c \leq c_{max}$, the index within cycle is $\varepsilon_c$ where, $1 \leq \varepsilon_c \leq m$.

$$y_{ij}^t \leq G_i^c Q_i^t, \ \forall i \in \mathcal{C}_I, \ \forall c \in \mathcal{T}, \ \forall t \in \mathcal{T}. \quad (10)$$

In the above equation, $G_i^c$ is the green split for the cell i at cycle-start-index $c$. The constraint is the restricted version of the CTM flow constraint. The $G_i^{\left\lfloor \frac{t-1}{m}+1 \right\rfloor}$ ensures the green split for the specific phase remains same throughout the cycle. The time index $t$ has relation with $c$ and $m$ as follows,

$$t = (c-1)m + \varepsilon_c \quad \forall t \in \mathcal{T} \quad \forall c \in \mathcal{T} \quad \forall i \in \mathcal{C}_I.$$

Total green split for any phase should not get below a minimum green split. The equation (11) assures that.

$$G_i^c \geq G_{min}, \ \forall i \in \mathcal{C}_I, \ \forall t \in \mathcal{T}, \ \forall c \in \mathcal{T}, \ \forall i \in \mathcal{C}_I. \quad (11)$$

Equation (12) ensures that one of the phase pairs of (1, 5), (2, 6), (3, 7), and (4, 8) (National Electrical Manufacturers Association (NEMA) convention) is allowed to move traffic from their subsequent cells. The presented model is can be applied to any signal control convention though.

$$G_{i+p}^c = G_{i+p+4}^c, \ \forall i \in \mathcal{I}, \ p \in \{1,2,3,4\}, \ \forall t \in \mathcal{T}, \forall c \in \mathcal{T}. \quad (12)$$

Similarly, the equation (13) assures there is synchronization between right and through turn green times. It makes sure one of the combinations (2, 9), (4, 10), (6, 11), or (8, 12) (NEMA convention) are allowed at a time if there is right turn allowed and no conflicts occur.

$$G_{i+2p}^c = G_{i+p+8}^c, \ \forall i \in \mathcal{I}, \ p \in \{1,2,3,4\}, \ \forall t \in \mathcal{T}, \ \forall c \in \mathcal{T}. \quad (13)$$

Finally, equation (14) makes sure that all the green fractions sum up to 1.

$$\sum_{p=1}^{4} G_{i+p}^{c} \leqslant 1, \quad \forall t \in \mathcal{T}, \quad \forall c \in \mathcal{T}, \quad \forall i \in \mathcal{I}. \tag{14}$$

### 2.3.2 Cycle-length Same as Discrete Time-interval (CSDT) Signal Control

The CSDT model is the only continuous-linear signal control model as mentioned earlier and was first discussed by Daganzo (1995). The model is presented in Eq. (15) which splits ($\omega_i^t$) the discrete time interval among all the phases and controls traffic by changing the saturation flow of the link. Detail formulation of the model that includes definition of the phases can be found in Ukkusuri, Ramadurai and Patil (2010). However, this model imposes trade-off between cell length and cycle-length and compromises accuracy and usability of the solution.

$$y_{ij}^{t} \leq \omega_i^{t} Q_i^{t}, i \in \mathcal{C}_I, j \in \Gamma(i), \forall (i,j) \in \mathcal{E}_{IM}, \forall t \in \mathcal{T} \tag{15}$$

### 2.3.3 Mixed-integer Traffic Signal Control (MISC)

For the purpose of comparing the performance of the SCRC model, a mixed integer signal control model is also implemented. The mixed integer signal control was presented by Aziz and Ukkusuri (2011). The implemented model is outlined below. Equation (16) constraints flow propagation where $\kappa_{ij}^t$ is a mixed-integer variable and can only be 0 or 1. $\kappa_{ij}^t = 1$, if the traffic movement is allowed at time $t$.

$$y_{ij}^{t} \leq \kappa_{ij}^{t} Q_i^{t} \quad \forall t \in \mathcal{T}, \quad \forall j \in \Gamma(i). \tag{16}$$

For any intersection at any time interval only one phase should be active. This can be assured by Eq. (17) where $\xi_p^t$ is another mixed-integer variable and activates one phase at a time.

$$\sum_{p \in P} \xi_p^t = 1 , \quad \forall t \in \mathcal{T}. \tag{17}$$

To integrate flow propagation with active phase, we can derive a relation between $\xi_p^t$ and $\kappa_{ij}^t$ using the following equation.

$$\kappa_{ij}^{t} = \sum_{p \in \sigma_{ij}} \xi_p^{t}, \quad \sigma_{ij} \subset P, \quad \forall t \in \mathcal{T}. \tag{18}$$

In the Eq. (18), $\sigma_{ij}$ is the number of phases that uses the cell connector at the intersection merge cell $i$ when $j$ is the successor ordinary cell. However, at a time only one phase is active. Hence, only one movement uses the cell at a time.

### 3. NUMERICAL RESULTS

In this section, System Optimal Dynamic Traffic Assignment with Signal Control (SO-DTA-SC) solution for a single destination network is discussed.

### 3.1 The Network Topology

The solved single destination network for System Optimal Dynamic Traffic Assignment with Signal Control (SODTA- SC) is presented in Fig. 1. In the figure, we can see that the network includes thirty five cells with two source cells (R1 and R2) and one destination cell (cell S). Each of the cells has length of 152.4 meters. There are two intersections in the network (node 3 and 4). The signal control models are applied to the cells 102, 202, 109, 209, 104, 204, 107, and 207. The intersection cells form the basis for through and right movements at the intersection. The physical parameters of all the cells and phase definitions are presented in Table. 1. The discrete time interval is t = 10 seconds and the assignment period is 110 time slots. The cycle-length for the SCRC was set to 6 time slots. For the sake of comparison of the results, we have ignored the minimum green time constraint that was presented in the formulation.

### 3.2 Signal Control Performance

- *It can be seen from our results that the SCRC and the CSDT model are more adaptive to the different traffic conditions and attain better objective-gain than the MISC. The SCRC has very marginal difference with CSDT in terms of objective value. The CSDT model attained the best objective-gain because it has flexibility to make the signal control decisions at every time slots. However, for small time slots (e.g. 10 sec) the cycle-length of the CSDT model is not realistic.*

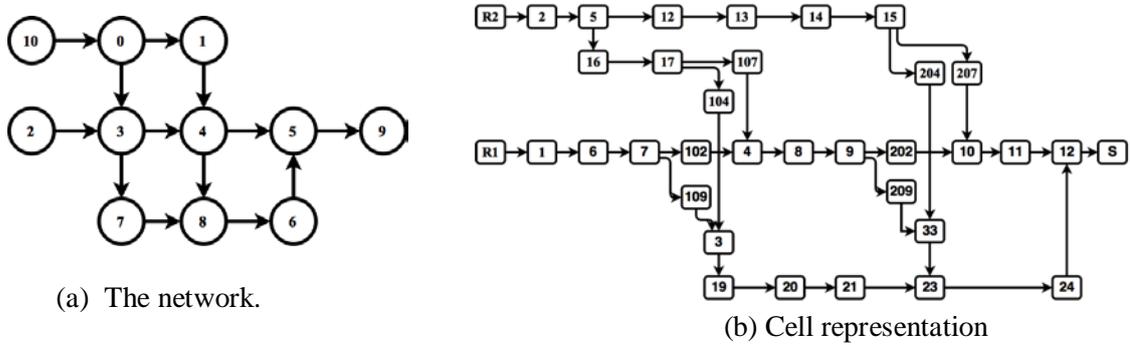

(a) The network.

(b) Cell representation

Figure 1. The network at the link level and cell level.

Table 1. The time invariant physical parameters of the cells

| Cells | 1,6,7,2,5,11 | All other cells | R1, R2 | S |
|---|---|---|---|---|
| $Q_i$ | 12 | 6 | 12 | 1 |
| $N_i$ | 36 | 18 | 1 | 1 |
| $\delta_i$ | 0.5 | 0.5 | -- | -- |
| v | 15.24 ms-1 | 15.24 ms-1 | -- | -- |

Only two phases are considered to produce results; phase-1: movements allowed from cells 102, 109, 202,209 and phase-2: movements allowed from cells 104, 107, 204, 207. However, the number of permitted phases can be easily extended to four or more. Results for all the three signal control models are generated using the same demand. They are tested for the objectives presented in Eq. (1), (2), and (3). Fig. 2b presents signal control green splits for the CSDT, MISC, and SCRC model where we can see that the SCRC (60 seconds of cycle-length) and CSDT models are flexible can take any positive value <= 1. This is due to the continuous property of both of the signal control methods. However, the MISC splits are either 0 or 1. Hence, SCRC and CSDT offer more flexibility to the optimization and are more adaptive to the traffic conditions.

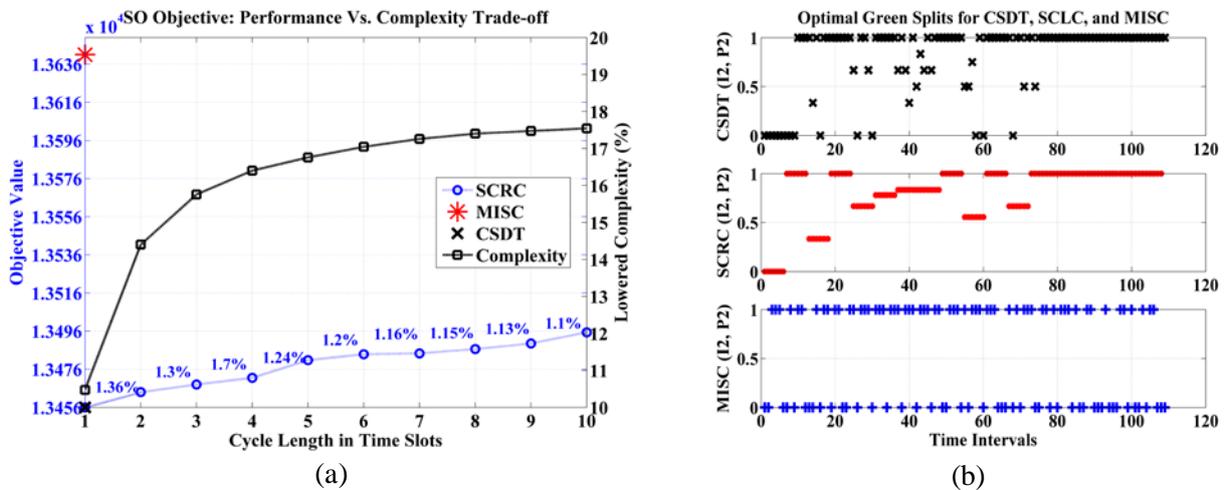

(a)

(b)

Figure 2. (a) Both the objective gain (shown in %) and reduction in complexity is computed relative to the MISC. The complexity is computed by adding the number of decision variables and constraints. (b) Phase-2 green splits for the intersection 2 (I2).

### 3.3 Performance Vs. Complexity Trade-off

- *We show that, regardless of the signal control model, for increasing accuracy, the complexity of the problem increases. Among the three, MISC has the highest complexity and due to the least flexibility in the choice of the green splits, it attains the lowest performance. The proposed SCRC model can significantly reduce the complexity. Just for the two intersections in the given network, the SCRC 10 time slot cycle-length (100 sec) reduces the complexity by 17.5%. For larger network, it is expected to improve the complexity gain even further.*

In Table. 2, a performance and complexity study is outlined where we can see that the CSDT model attains better objective values for all the three objective functions. However, this model has higher complexity than the SCRC model. For the 60 seconds of cycle-length, SCRC attains comparable objective value and substantially less complexity. For a practically large network, the SCRC approach to signal control is expected to provide significant gain in complexity at a nominal reduction in the gain in objective value. This trend is already indicative in Fig. 2a for the two-intersection network under study.

- *We have found that for increasing cycle-length of the SCRC model, the objective-gain degrades marginally, though the complexity reduces substantially (Fig. 2a). This model also enables us to set a realistic cycle-length without compromising the discrete time interval of the DNL models.*

Table 2. Performance and complexity comparison for the different signal control schemes

| Signal control | Objective | Time to solve (sec) | Variables | Constraints | Objective Value |
|---|---|---|---|---|---|
| CSDT | SO | 2.35 | 9105 | 20982 | 13456 |
| | Flow benefit | 2.5 | | | 13401.705 |
| | DCS | 2.4 | | | 15777.067 |
| MISC | SO | 83.01 | 9993 | 23614 | 13641 |
| | Flow benefit | 243.938 | | | 13586.866 |
| | DCS | 106.3 | | | 16002.422 |
| SCRC (6slots) | SO | 2.33 | 8369 | 19510 | 13483.857 |
| | Flow benefit | 2.259 | | | 13429.517 |
| | DCS | 2.3 | | | 15805.107 |

In Fig. 2a, we can see that the CSDT, and SCRC with 1 slot cycle-length attains same objective value. For a practical choice of cycle-length (e.g. 7 slots/70 sec), complexity and objective values are improved by 17.25% and 1.15% respectively. Hence, this model can be applied to solve any large network and able to attain solution faster than the MISC model.

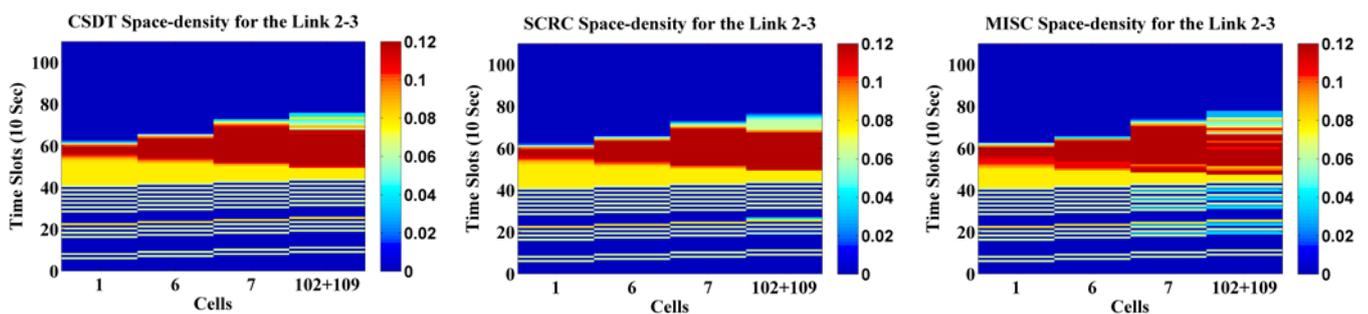

Figure 3. Space-density plot when the objective is set to flow benefit.

### 3.4 Traffic Evolution

- *It is noticeable from the results that continuous-linear signal control models (i.e. CSDT, SCRC) are more adaptive to the queue formations at the intersections and traffic evolution on the links. As a result, they are more effective for reducing the congestion on the links and more resilient against different traffic conditions.*

In Fig. 3, the density on the link 2-3 is presented for all the three signal control models using the flow benefit objective defined in Eq. (2). For the first 40 time slots, very light on-off demands were generated. As result, very light density is visible during that time period. After 40th time slot, moderate to very high demands were generated. As a consequence, we can see highly congested traffic condition and shock-waves. The SCRC and CSDT signal control attain similar traffic evolution over the link. On the other hand, due to the binary nature of the MISC, which is less flexible, we can see more congestion throughout the link and also at the intersection.

## 4. SUMMARY

In this paper, we propose a novel traffic signal control model, namely, Signal Control with Realistic Cycle-length (SCRC) that can resolve the trade-off between cell-length and signal control cycle-length. The key concept of using two time scales can be applied in conjunction with any discrete time Dynamic Network Loading (DNL) model (e.g. TTM, Link Transmission Model, etc.) to formulate a continuous and linear SO-DTA-SC optimization problem.

We have found that the proposed SCRC model attains objective values comparable to that of the CSDT scheme. The advantages of the presented SCRC are, it has linear continuous formulation, is more adaptive to varying traffic conditions, and more effective to reduce congestion. In addition to using more realistic cycle lengths than that with CSDT, this approach induces less decision variables, hence, reduces the complexity of the problem substantially. As a result, this model can attain optimal solution faster than the CSDT or the MISC model. In future, we will investigate the impact of signal control on the emission using our SO-DTA-SC framework.